\documentclass[useAMS,usenatbib]{mn2e}

\usepackage{graphicx}
\usepackage{amssymb}
\usepackage{amsmath}
\usepackage{tabularx}
\usepackage{color}
\usepackage{subfigure}

\title{Finding New High-Redshift Quasars by Asking the Neighbours}
\author[K. L. Polsterer, P.-C. Zinn and F. Gieseke]{
Kai Lars Polsterer$^{1}$\thanks{E-mail:polsterer@astro.rub.de},
Peter-Christian Zinn$^{1,2}$\thanks{E-mail:zinn@astro.rub.de} and
Fabian Gieseke$^{3}$\thanks{E-mail:f.gieseke@uni-oldenburg.de}\\
$^{1}$Ruhr-University Bochum, Faculty of Physics and Astronomy, Universit\"atstra{\ss}e 150, 44801 Bochum, Germany\\
$^{2}$CSIRO Astronomy \& Space Science, PO Box 76, Epping, NSW, 1710, Australia\\
$^{3}$Carl von Ossietzky University, Department of Computer Science, Uhlhornsweg 84, 26129 Oldenburg, Germany}
\begin{document}

\date{Accepted 2012 September 17. Received 2012 September 14; in original form
2012 April 26}

\pagerange{\pageref{firstpage}--\pageref{lastpage}} \pubyear{2012}

\maketitle

\label{firstpage}

\begin{abstract}
Quasars with a high redshift (z) are important to understand the
evolution processes of galaxies in the early universe. However only a few of
these distant objects are known to this date. The costs of building and
operating a \mbox{10\,-metre} class telescope limit the number of facilities
and, thus, the available observation time. Therefore an efficient selection of
candidates is mandatory. This paper presents a new approach to select quasar candidates with
high redshift ($\rm{z}>4.8$) based on photometric catalogues. We have chosen to
use the $\rm{z}>4.8$ limit for our approach because the dominant Lyman\,$\alpha$
emission line of a quasar can only be found in the Sloan $i$ and $z$-band
filters. As part of the candidate selection approach, a photometric redshift
estimator is presented, too. Three of the 120,000 generated candidates have been
spectroscopically analysed in follow-up observations and a new $\rm{z}=5.0$
quasar was found. This result is consistent with the estimated detection ratio
of about 50 per cent and we expect 60,000 high-redshift quasars to be part
of our candidate sample. The created candidates are available for
download at MNRAS or at {\tt
http://www.astro.rub.de/polsterer/quasar-candidates.csv}.
\end{abstract}

\begin{keywords}
methods: statistical, techniques: photometric, galaxies: distances and
redshifts, quasars: general, catalogs, surveys
\end{keywords}

\section{Introduction}

In the late 1950's, observations in the radio regime discovered quasi-stellar
radio sources with no optical counterpart. Later, a quasar was
found to be a distant galaxy with an active galactic nucleus ({\it AGN}).
An {\it AGN} is a central super-massive black hole which accretes material
and thereby creates radiation. This phenomenon is seen in different ways and
therefore creates a large number of observable {\it AGN} classes
\citep[][]{urry:1995}. \citet{anto:1993} uniformly describes this zoo of
{\it AGN}s.

Due to their extreme luminosity, even very distant quasars can be observed. This
allows to study processes in the early universe. Another benefit of their
intrinsically high luminosity is the possibility to find these sources even in
surveys with low detection levels. A representative sample of high-z quasars
would help to understand the formation process of galaxies \citep[][]{whit:1991}
and the influence of super-massive black holes on galaxy evolution
\citep[][]{catt:2009}. The formation of larger galaxies through hierarchical
clustering has direct effects on the creation of quasars. \cite{carl:1990} found
that the birth-rate of quasars is proportional to the rate of mergers of
gas-rich galaxies. The presence of an {\it AGN} has direct consequences for the
hosting galaxy. As soon as the {\it AGN} starts to accrete material and to
produce radiation, the gas content of the bulge is heated and, in dependence on
the strength of the radiation, blown away. This directly leads to a stop of star
formation in the bulge \citep[][]{catt:2009}. Currently there are only a few
quasars known with a redshift of $\rm{z}>4.8$. Therefore statistics on their
number density is not reliable \citep{chri:2004}.

As larger optical telescopes tend to improve both sensitivity and spatial
resolution, the field of view is reduced correspondingly. A reduced field of
view requires an appropriate selection of targets based on previous observations.
Today, large panoramic catalogues are available which can be used for the target
selection process.

The presented method of photometrically selecting high-redshift quasar
candidates is based on the Sloan Digital Sky Survey ({\it SDSS}) {\it DR6}
\citep[][]{york:2000,adel:2008}. A dedicated 2.5\,-metre telescope at the Apache
Point Observatory in New Mexico was used to create an imaging and spectroscopic
catalogue of the northern Galactic Cap ($9,583\,\rm{deg}^2$). Images have been
taken in drift-scan mode using five broadband filters ($u$, $g$, $r$, $i$, $z$:
300\,nm to 1,000\,nm). For a selected sub-sample a spectroscopic analysis has
been conducted with a fibre-fed spectrograph. In total 287 million objects are
stored in the catalogue of which 1.27 million have associated spectra.

The increasing amount of data which is available in catalogues can no longer be
handled manually. It requires an automated processing and selection of
scientifically important objects. The spectroscopically observed objects have
been automatically classified by correlating them with 33 template spectra. In
\citet{schn:2010} a hand-vetted catalogue of 105,783 spectroscopically confirmed
quasars is presented. This catalogue contains 1,248 objects with $\rm{z}>4.0$
whereof 56 objects have a redshift of $\rm{z}>5.0$. It is used as a reference
set for the presented photometric selection method.

Throughout this paper we adopt a flat $\Lambda$CDM cosmology with
$H_0= 70.2\,{\rm km\,s^{-1}\,Mpc^{-1}}$ and $\Omega_\Lambda = 0.727$
\citep[][]{koma:2011}.

\paragraph*{Outline:}
The presented technique to select quasar candidates is realised by
combining a redshift estimation and a classification approach. In \mbox{Section
\ref{redshiftEstimation}} the redshift estimator and the used reference sample
are presented. The classification approach and the way we compose the
multi-stage classifier are described in \mbox{Section \ref{selection}}. In
\mbox{Section \ref{results}} the generated candidate sample is presented (along
with theoretical number expectations), and the results of follow-up observations
are discussed. \mbox{Section \ref{conclusions}} concludes this work.

\section{Photometric Redshift Estimation}
\label{redshiftEstimation}
 
\begin{figure}
\begin{centering}
\includegraphics[width=1.0\columnwidth]{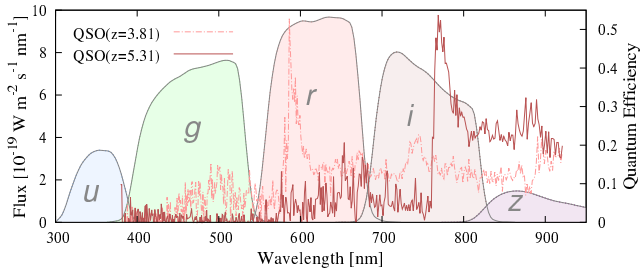}
\caption{Spectra of the high-z quasar {\it SDSS} J082547.79+332836.9 ($\rm{z}=3.81$)
  and {\it SDSS} J165902.12+270935.1 ($\rm{z}=5.31$). In the background the wavelength
  dependent quantum-efficiency of the $u$, $g$, $r$, $i$, $z$-bands of the {\it
  SDSS} are plotted. The Lyman\,$\alpha$ emission line of the quasar is red-shifted to the
  $r$ and $i$-band.}
  \label{sdss_comparison}
\end{centering}
\end{figure}

An efficient creation of a high-z quasar candidate catalogue requires a
precise photometric estimation of the redshift. These estimates allow the
rejection of candidates with a redshift below a certain threshold.
Photometric methods try to detect the position of strong emission and absorption
features which are unresolved in the broadband filters. In \mbox{Figure
\ref{sdss_comparison}} the spectra of two high-z quasars are presented
together with the filter bands of {\it SDSS}. The Lyman\,$\alpha$ emission lines of
these quasars dominate the flux in the $r$ and $i$-band, respectively.

\subsection{Related Work}

Several methods for photometric redshift estimation can be found in the
literature. These estimation methods either rely on physical assumptions or on
the quality of the considered reference set.\footnote{With increasing size of
the reference set the estimation quality increases while the processing speed
drops for most methods.} We will briefly discuss these two concepts. 

\subsubsection{Template Fitting}

A classical way of photometric redshift estimation is the fitting of spectral
energy distributions ({\it SED}s). This method is typically limited to a small
set of representative model spectra. These spectra can be created based on empirical or
simulated {\it SED}s. In \citet{bolz:2000}, an estimation method using a
$\chi^2$-minimisation is presented. The quality of the fit is highly dependent
on the applied template spectra. When the used photometric features do not allow to
distinguish between the different {\it SED}s, the fit can create
catastrophic outliers. The big advantage of the {\it SED} fitting approach is the
ability to predict rough estimates even for objects with spectroscopically yet
unobserved redshifts.

\citet{wu:2010} present a photometric redshift estimator that combines
{\it SDSS} and {\it UKIRT} Infrared Deep Skey Survey ({\it UKIDSS}) data. A
reference sample of 7,400 quasars with $0.5<\rm{z}<5.2$ was divided into 91
redshift bins. Each bin was analysed and a median colour was calculated for the
eight directly neighbouring bands. The most probable redshift is retrieved by
applying a $\chi^2$-minimisation of colours with respect to the photometric
errors. As a result of their tests, 71.8 per cent of their reference sample have
a redshift estimation error of $|\rm{z_{spec}-z_{est}}|<0.1$.

\subsubsection{Neural Networks}

Another way of redshift estimation is based on empirical reference data. In
\citet{mill:2011}, such an approach is presented for the {\it SDSS} data.
Here, 80 per cent of the main galaxy sample \citep[][]{stra:2002}, 10 per cent
of the luminous red galaxy sample \citep[][]{eise:2001}, and 10 per cent of the
active galactic nuclei sample \citep[][]{kauf:2003} have been combined to a set of 550,000
objects with spectroscopic redshifts. Half of this reference set was used for
training an artificial neural network and the other half for testing. The input
nodes represent the {\it SDSS} magnitudes, the concentration index and the Petrosian
radii in the $g$ and $r$-band. Three hidden layers, of 14 neurons each, have
been used to calculate the redshift output. They limited their tests to a redshift
range of $\rm{z}<0.4$. In this range their estimates deviate from the
spectroscopically determined redshifts with an $\rm{rms}\approx 0.03$.

\cite{laur:2011} consider combining a clustering approach with neural networks.
To build the model, the data is clustered in the feature space. These clusters
are used to train individual neural networks and, thus, to obtain a more
flexible regression model.

\subsection{k-Nearest Neighbours}

In \mbox{Section \ref{selection}}, we present a new quasar selection method.
Fundamental for this selection is a reliable photometric redshift estimation.
A new redshift estimator was developed which is based on empirical data. It was
designed to process large reference sets without requiring physical assumptions
or models. This estimator realises the important step of rejecting candidates
with low redshifts. It uses a k-nearest neighbours ({\it kNN}) regression model
to predict redshifts \citep[][]{hast:2009}.\footnote{This is similar to the approach
presented in \citet{csab:2003} to estimate redshifts of galaxies with
$\rm{z}\leq 0.5$.}

\subsubsection{Regression Model}
Given the reference sample $S = \{ ({\bf x}_1,y_1), \ldots, ({\bf x}_n,y_n)\}
\subset {\mathbb R}^d \times {\mathbb R}$, the predicted redshifts $\hat Y$ are
calculated from the redshift values $y_i$ of the $k$ Euclidean closest objects.
Thereby the neighbourhood $N_k({\bf x})$ is determined on basis of the
representation of the reference objects ${\bf x}_i$ in the
feature space:

\begin{equation}
 \hat Y({\bf x})=\frac{1}{k}\sum \limits_{{\bf x}_i \epsilon
 N_k({\bf x})} y_i
\end{equation}

To retrieve these $k$ neighbours $N_k({\bf x})$ of
${\bf x}$ efficiently, a $k$-d-tree is used \citep{bent:1975}. This data
structure is a binary search tree that allows a spacial look-up in ${\mathcal O}
(\log_2 n)$ instead of ${\mathcal O} (n)$ time. Besides the redshift value the
standard deviation of the $k$ nearest $y_i$ is calculated as a quality measure. High standard
deviations indicate a bad coverage of the target space. To analyse the
distribution of reference objects in the feature space, the length of the
average distance vector to the $k$ nearest neighbours can be calculated. Large
values indicate that the requested object lies in a sparsely populated region of
the feature space and might therefore have a very high or low redshift,
respectively.\footnote{The disadvantage of the considered {\it kNN} regression
model is its limitation to predict only values that are covered by the reference sample.}

\begin{table}
\begin{centering}
  \begin{tabular}{c|c|c}
  {\it SDSS} PhotoObjID & z & Reference \\\hline\hline
  588023045868553340&5.79&\citet{fan:2006}\\\hline
  587740525079167786&5.80&\citet{fan:2004}\\\hline
  587727942951109703&5.82&\citet{fan:2001}\\\hline
  587733411521299389&5.83&\citet{fan:2006}\\\hline
  587731186204541926&5.85&\citet{fan:2004}\\\hline
  587737808499572747&5.85&\citet{fan:2006}\\\hline
  587736914601902923&5.93&\citet{fan:2004}\\\hline
  587738951494075293&5.93&\citet{fan:2006}\\\hline
  587729157893456734&5.99&\citet{fan:2001}\\\hline
  587741421098303812&6.00&\citet{fan:2006}\\\hline
  587738615416554219&6.01&\citet{fan:2006}\\\hline
  587729751132603659&6.05&\citet{fan:2003}\\\hline
  587735666926158831&6.07&\citet{fan:2004}\\\hline
  587739608093491422&6.13&\citet{fan:2006}\\\hline
  587736783608677304&6.22&\citet{fan:2004}\\\hline
  587732482206139341&6.23&\citet{fan:2003}\\\hline
  587728881415553909&6.28&\citet{fan:2001}\\\hline
  588013383815791587&6.43&\citet{fan:2003}
  \end{tabular}
  \caption{High-z quasars which have been found based on the $i$-band dropout
  method and that have been used to extend the reference sample.}
  \label{highzExtension}
\end{centering}
\end{table}

\subsubsection{Reference Sample}
\label{redshiftReferenceSample}

A reference sample was created to support the detection process of high-redshift
quasars, which is based on the {\it SDSS} quasar catalogue
\citep[][]{schn:2010}. This sample is used to populate the feature space. To
increase the processing speed of estimating redshifts, the size of the reference
sample was reduced. For this reason the input catalogue was split into three
subsets:

\begin{enumerate}
  \item The low-redshift ($\rm{z}<2.0$) set with 81,238, 
  \item the medium-redshift ($2.0\leq \rm{z}< 4.0$) set with 22,696,
  \item and the high-redshift ($\rm{z}\geq 4.0$) set with 1,258 objects.
\end{enumerate}

As the high-redshift set contains only a few quasars with $\rm{z} \geq 5.0$ and
is limited to $\rm{z_{max}} = 5.5$, 18 additional objects with {\it SDSS}
features have been added (see \mbox{Table \ref{highzExtension}}). To create a
homogeneously distributed sample (what we call the reduced sample),
all quasars have been assigned to 120 equal bins in redshift space for redshifts between $\rm{z}=1.0$ and
$\rm{z}=7.0$. We deliberately excluded all quasars below ${\rm z}=1.0$
since the reference sample is specifically designed for finding high-z quasars.
Hence, the many quasars below ${\rm z}=1.0$ would just slow down processing
without substantially adding information with respect to the anticipated redshift
range. For those bins below $\rm{z}=4.8$, the size was limited to ten reference
objects and supernumerous objects have been randomly extracted. Above a redshift
value of $\rm{z}=4.8$, all quasars have been included. Due to missing high-z
references the high-z bins are not filled equally. The resulting reference
sample contains 1,106 quasars with spectroscopically determined redshifts and
{\it SDSS} magnitudes.

\subsubsection{Model Selection}

During tests with the {\it kNN} regression model it turned out that the best results
are achieved when using colours ($u-g$, $g-r$, $r-i$, $i-z$) instead
of magnitudes. This may be caused by distribution effects of the reference
objects in the Euclidean feature space which are induced by intrinsic object
characteristics like their luminosity. These effects are minimised by a kind of
normalisation which is obtained by the dimension reduction from filter band to
colour space. Colours eliminate the effects of different luminosities
because they just reflect differences in flux density and not the flux density
itself. Therefore we gain predictive quality by treating faint and bright
sources with same redshifts as equal. The $k$-value was set to eight, based on
tests of the redshift estimation performance. With smaller $k$-values the
standard deviation of the estimation errors increases. Slightly larger values
(up to 20) did not significantly improve the results but enhanced the
computational effort.

\subsection{Evaluation of the Redshift Estimation}

The quality of the {\it kNN} regression redshift estimation approach was tested
on the quasars with spectroscopic redshifts taken from \citet{schn:2010}. As the
reference sample is limited to redshifts with $\rm{z} \geq 1.0$, quasars with
lower redshifts have been excluded from the tests. There are only a few high
redshift quasars known. Therefore all objects are required as reference as
well as for testing.\footnote{To prevent any bias from objects that are part of both
the test and reference sample, objects are not considered as reference when their
value is estimated for testing.} 

\begin{figure}
\begin{centering}
\includegraphics[width=1.0\columnwidth]{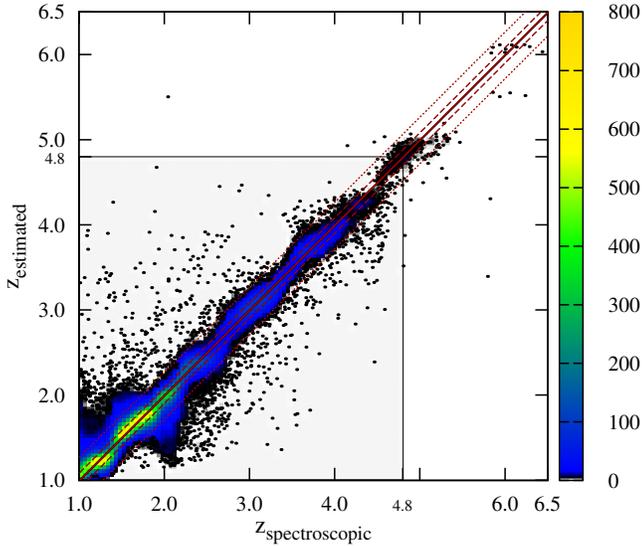}
\caption{Comparison of estimated and spectroscopic redshifts where the
  estimation is based on a reference sample consisting of all known quasars
  above $\rm{z}=1.0$. The mean value, the $1\sigma$- and the $3\sigma$-values of
  the fitted Gaussian distribution are plotted as a solid, a dashed and a
  dotted line, respectively. In addition the interesting area of redshifts with
  $\rm{z}>4.8$ is marked.}
  \label{estimationQuality1}
\end{centering}
\end{figure}

\begin{figure}
\begin{centering}
\includegraphics[width=1.0\columnwidth]{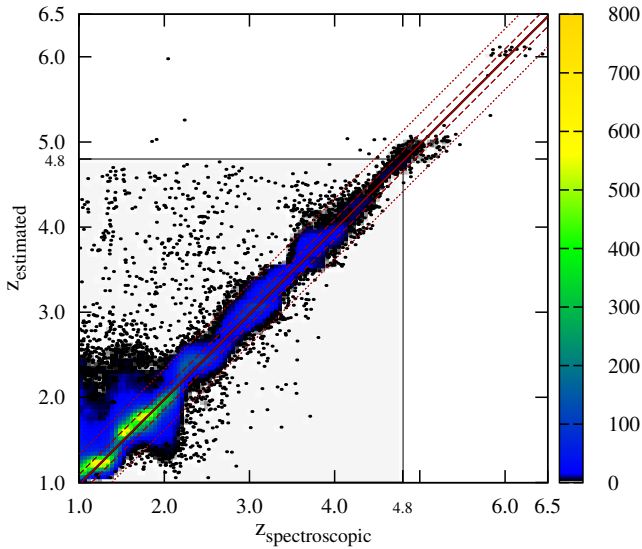}
\caption{Results of the estimation done with the reduced reference set. The
considered $\rm{z}=4.8$ limit, as well as the parameter of the fitted Gaussian
distribution are marked, as for \mbox{Figure
{\ref{estimationQuality1}}}.}
  \label{estimationQuality2}
\end{centering}
\end{figure}

\begin{figure}
\begin{centering}
\includegraphics[width=0.85\columnwidth]{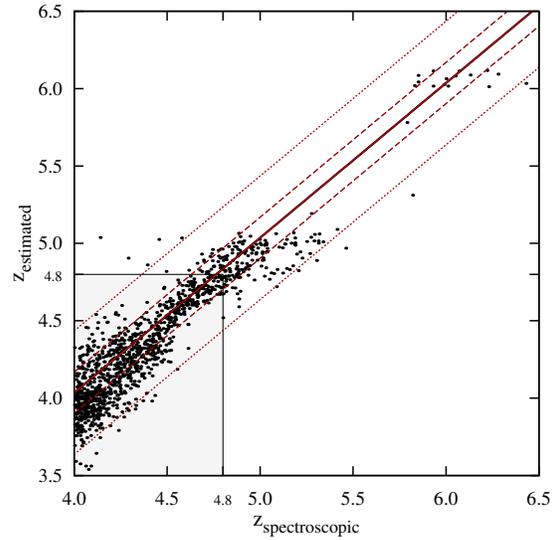}
\caption{Magnified high-z ($\rm{z}>4.0$) sector of the results of the estimation
done with the reduced reference set. Important parameters are marked in the same
manner as in the previous figures.}
  \label{estimationQuality3}
\end{centering}
\end{figure}

\subsubsection{Results}

A leave-one-out cross-validation scheme was applied to optain the test results.
For each object of the test sample, a deviation $\Delta \rm{z} =
\rm{z_{spectroscopy}} - \rm{z_{estimation}}$ and a redshift independent
deviation $\Delta \rm{z} / (1+\rm{z_{spectroscopy}})$ was determined. In
\mbox{Figure \ref{estimationQuality1}}, the results of the estimation approach
are presented. The area below $\rm{z}=4.8$ is marked grey because quasar
candidates which lie outside of the target redshift range are rejected. The
comparison of the results of the reduced reference sample (\mbox{Figure
\ref{estimationQuality2}}) with the results of the complete reference sample
(\mbox{Figure \ref{estimationQuality1}}) demonstrates that both sets perform
equally well for $\rm{z}>4.8$. In \mbox{Figure \ref{estimationQuality3}}, the
results of the important $\rm{z}>4.0$ area are shown in a magnified plot. With the
reduced reference sample lower red-shifted quasars can still be processed with
an appropriate estimation quality, even though the size of the reference set was
dramatically reduced from 77,096 to 1,106 objects. As the reduced reference
sample is homogeneously distributed in redshift, no catastrophic outliers are
produced by an over-representation of lower redshifts. In both \mbox{Figures
(\ref{estimationQuality1} and \ref{estimationQuality2})}, a redshift dependent
fluctuation between estimated and spectroscopic values can be observed. This is
directly connected to the passage of the Lyman\,$\alpha$ features through the
broadband filters.

\begin{figure*}
\begin{centering}
\subfigure[Estimation error of all known $\rm{z}\ge 1.0$ quasars with complete
reference sample.]{
\includegraphics[width=0.32\textwidth]{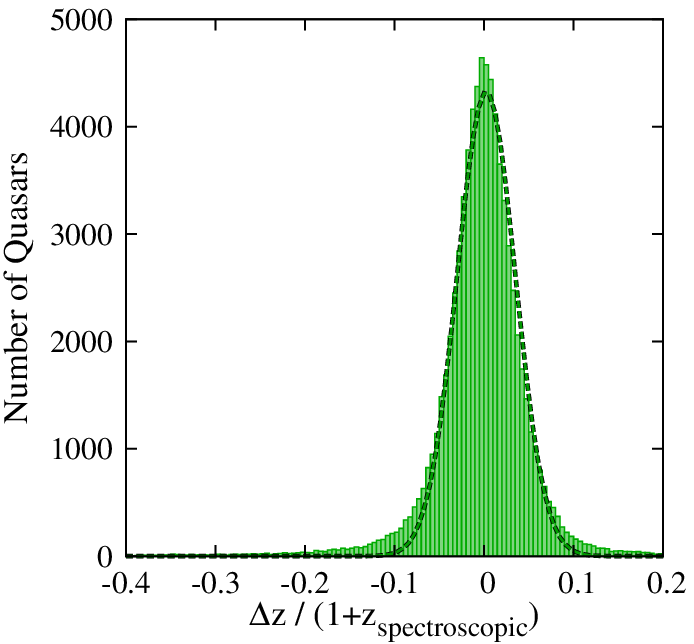}}
\hfill
\subfigure[Estimation error of all known $\rm{z}\ge 1.0$ quasars with reduced
reference sample.]{
\includegraphics[width=0.32\textwidth]{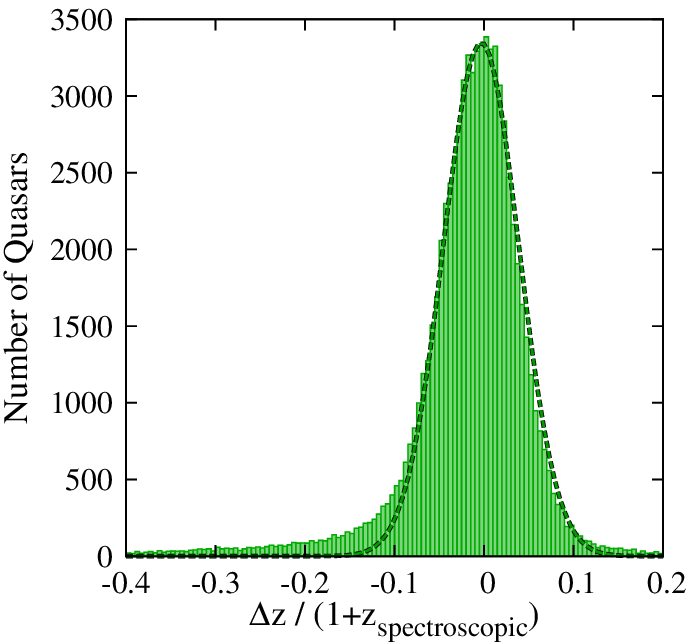}}
\hfill
\subfigure[Estimation error of all known $\rm{z}\ge 4.0$ quasars with reduced
reference sample.]{
\includegraphics[width=0.310\textwidth]{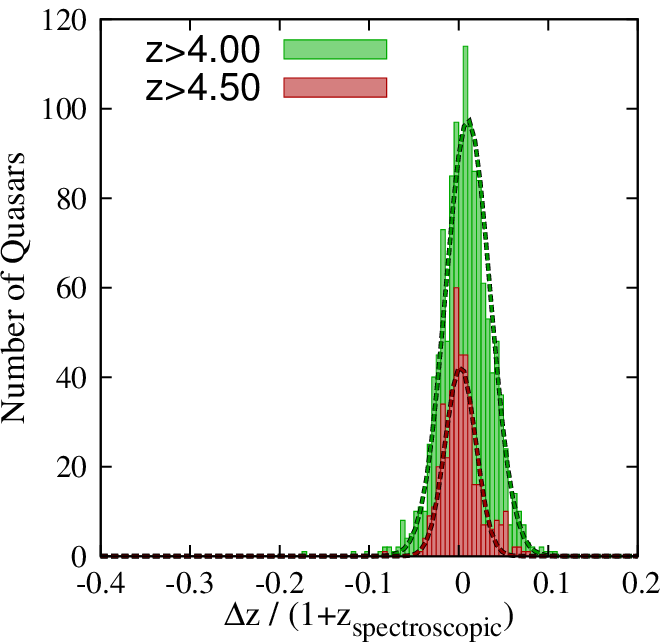}}
\caption{Histograms of the redshift estimation errors are compared for
  different reference and test samples. The fitted Gaussian distributions are
  plotted as double dashed lines.
  }
  \label{estimationQuality4}
\end{centering}
\end{figure*}

\begin{table*}
\begin{centering}
  \begin{tabular}{c|r|c|r|r|r|r|r}
  Reference & Reference Size & Redshift & Test Set Size &
  $\langle\frac{\Delta \rm{z}}{1+\rm{z}}\rangle$ & \hspace{3ex}$\sigma_\frac{\Delta \rm{z}}{1+\rm{z}}$ & $\langle{\Delta \rm{z}}\rangle$ &
  \hspace{3ex}$\sigma_{\Delta \rm{z}}$\\\hline\hline
  complete & 77,096 quasars & {$1.0 \le \rm{z} < 6.5$} & 77,096
  quasars & 0.003 & 0.033 & -0.007 & 0.091 \\\hline
  complete & 77,096 quasars & {$4.0 \le \rm{z} < 6.5$} & 1,258
  quasars & 0.008 & 0.024 & -0.065 & 0.126 \\\hline
  complete & 77,096 quasars & {$4.5 \le \rm{z} < 6.5$} & 406 quasars
  & 0.002 & 0.021 & -0.011 & 0.117 \\\hline
  reduced & 1,106 quasars & {$1.0 \le \rm{z} < 6.5$} & 77,096
  quasars & -0.023 & 0.095 & -0.024 & 0.119 \\\hline
  reduced & 1,106 quasars & {$4.0 \le \rm{z} < 6.5$} & 1,258 quasars
  & 0.010 & 0.025 & 0.036 & 0.133 \\\hline
  reduced & 1,106 quasars & {$4.5 \le \rm{z} < 6.5$} & 406 quasars &
  0.002 & 0.016 & 0.001 & 0.087
  \end{tabular}
  \caption{Results of fitting Gaussian distributions to the redshift
  estimation errors. The values are given for both $\Delta \rm{z}/(1+\rm{z})$
  and $\Delta \rm{z}$. Additionally, the errors have been determined on both the
  reduced and the complete reference sample.}
  \label{redshiftErrorFits}
\end{centering}
\end{table*}

The deviation $\Delta\rm{z}$ and the redshift-independent deviation
$\Delta \rm{z/(1+z)}$ was determined for different redshift ranges. The
results are presented in \mbox{Table \ref{redshiftErrorFits}}. The
distribution of the estimation errors for the full redshift range
$1.0<\rm{z}<6.5$ is presented in \mbox{Figures \ref{estimationQuality4}\,a and
\ref{estimationQuality4}\,b} for the full and the reduced reference sample,
respectively. In comparison to the tests on the full redshift range, better
results are achieved for the high-redshift quasar (see \mbox{Figure
\ref{estimationQuality4}\,c}). As it was intended, the reduced reference sample
performs better on quasars in the targeted selection range with $\rm{z}>4.8$.
This is caused by the better representation per redshift bin of $\rm{z}>4.8$
quasars in the feature space.

\subsubsection{Comparison}

The performance of the redshift estimation is comparable to the results of
\citet{wu:2010}, but: (i) Their results are based on a test sample of only 8,498
quasars which was partially (87 per cent) used to create the median colours.
This creates a bias on the test results. (ii) {\it SDSS} and {\it UKIDSS} data were
used. (iii) The presented results are dependent on the redshift range and can
therefore not be compared directly.

\citet{mort:2011} present a Bayesian redshift estimator which is used to
assign observation priorities to their high-redshift quasar candidates. Based on
photometric data of {\it SDSS} and {\it UKIDSS} an accuracy of
$|\rm{z_{spec}-z_{est}}|\simeq0.1$ is presented for a redshift range of $5.8 < \rm{z} < 7.2$.
This deviation is comparable to the results achieved with the presented {\it kNN} regression
model. As there are no quasars known with $\rm{z}> 6.5$ in both {\it SDSS}
and {\it UKIDSS}, the results have been computed with simulated {\it SED}s and
are based on a modelled high-z quasar population.

In \citet{card:2010}, a 32-band data set is used to calculate photometric
redshifts. They present a $1\sigma$ scatter in $\Delta \rm{z}/(1+\rm{z})$ of
0.008 for $0.1 < \rm{z} < 1.2$, 0.027 for $1.2 < \rm{z} < 3.7$ and 0.016 for
$\rm{z} > 3.7$, respectively. For the high-redshift range, these results are as
good as the results of the {\it kNN} regression model which uses only five
bands.

A direct comparison with \citet{laur:2011} indicates a competitive but
slightly better performance of our regression model.\footnote{A direct
comparison of the regression models is difficult due to different test samples.}
Creating global models for sub-clusters in the feature space is an improvement
compared to using pure global models. The {\it kNN} regression model, however,
creates local predictions that seem to reflect the sample intrinsic error rate.

The developed redshift estimator exhibits a competitive performance compared
to the other presented approaches. Its main benefit is the capability to predict
photometric redshifts with comparable quality even with less photometric bands.
Additionally the processing speed on large reference samples is increased by
using special data structures. The most important advantage of our approach is
that no physical assumptions are required at all.

\section{Photometric Selection of Quasars}
\label{selection}

The described estimation of redshifts is dependent on a reliable pre-selection
of quasars. The main problem in selecting these candidates is that broadband
observations of the high-redshift {\it SED}s of quasars become similar to
the {\it SED}s of cool stars. Before describing the {\it kNN}-based quasar selection
approach and the creation of the required reference samples, an overview of
currently available methods is given.

\subsection{Related Work}

Several methods are applied to find high-redshift quasar candidates. Among these
methods are linear and Bayesian models, which we describe next.

\subsubsection{Linear Models}

In the {\it SDSS}, the selection of quasar candidates for spectroscopic observations
is based on photometric data \citep[][]{rich:2002}. The magnitudes which are
extracted on basis of fits to a point spread function ({\it PSF}) are inspected for
unresolved objects in distinct 3D colour spaces to separate quasars from
stars. Several decision trees have been created that reflect relations in
band-flux and thereby define regions in the feature space.

The $\rm{z}>5.8$ quasars that are presented in
\citet{fan:2001,fan:2003,fan:2004,fan:2006} have been detected by using an $i$-band
dropout technique in combination with {\it 2MASS} magnitudes. This technique
assumes no detection in the $u$, $g$, $r$-band, a weak detection in the
$i$-band, and a detection in the $z$-band. The principle behind this approach is
that the strong Lyman\,$\alpha$ forest absorption enters the $i$-band at
$\rm{z}>5.5$. The resulting constraints are: $mag_z<20.2$ with $\Delta
mag_z<0.1$, $mag_i-mag_z > 2.2$ and $mag_z - mag_J < 1.5$. This method turned
out to have a high false positive rate. Only about $3$ per cent of the
candidates that had follow-up observations were confirmed as quasars.

\citet{wu:2010} present a quasar selection approach that uses colour-colour
relations in {\it SDSS} and {\it UKIDSS} bands. The best solution to separate
quasars from stars was empirically found in the $Y-K$ and $g-z$-colour space
with the linear relation $Y-K > 0.46 (g-z) + 0.53$. This relation correctly
separates 97.7 per cent of both the 8,498 quasars and 8,996 stars which have
been used as reference. Unfortunately this simple linear separation fails for quasars with
$\rm{z}>4.0$. For this reason the relation $J-K > 0.45 (i-Y) + 0.64$ has
been created to the find 99 per cent of the 101 reference quasars with
$\rm{z}>4.0$. The downside of this framework is that the contamination with stars is more than
doubled. Another problem of optimising the separating relation is that the equal
cardinalities of the reference samples do not necessarily reflect the real
distribution. Even at larger distance from the galactic plane the number density
of cool stars exceeds the number density of high-redshift quasars by a factor of
up to 1,000 \citep[][]{fan:1999}.

\subsubsection{Bayesian Models}

\citet{mort:2011} present a probabilistic candidate selection approach which
uses {\it SDSS} and {\it UKIDSS} to find the most probable high-z quasars. Their
approach is comparable to their redshift estimator. It uses a Bayesian model to
separate quasars from stars. Their targeted redshift range is $\rm{z} > 5.8$.
With their approach a reduction of the primary data set by a factor of
about $2.5\times10^4$ to 893 candidates was realised. Thereby the
probabilistic evaluation of each object took 0.01\,s to 0.1\,s. Only 88 of these
893 candidates turned out to be real detections of astronomical sources with three
previously known high-z quasars. Follow-up observations left seven photometric
candidates of which four have an estimated redshift $\rm{z}\simeq 6.0$. The
results of these candidates are not published yet.

\subsection{Quasar Selection with a kNN Classifier}
\label{kNNquasar}

The main purpose of the quasar selection scheme presented below is to create
quasar candidates with a high probability for follow-up observations. For this reason
the number of recovered quasars is set to a lower percentage than in the other
presented approaches.

\subsubsection{Pre-Processing}
As a pre-filtering of all 287 million {\it SDSS} objects a limiting $i$-band magnitude
of 16.5\,mag is used. Brighter objects or objects with an $i$-band error
$>0.2$\,mag are ignored, since at our redshift cut of $\rm{z}>4.8$, an $i$-band
magnitude of 16.5\,mag corresponds to a restframe (at 1,450\,$\rm \AA$) absolute magnitude of
$M_{1450}=-32$. Additionally, only point-like or slightly extended objects are
selected. To separate point sources from extended ones {\it SDSS} uses the {\it PSF}
and model magnitudes: When an object complies with $\rm{PSF_{mag}-model_{mag}} > 0.145$
in at least two of the three $g$, $r$, $i$-bands, it is labelled as
galaxy in the {\it SDSS} catalogue. Here, a value of 0.3 is used instead to include
slightly extended sources of e.g., possible lensed quasars. The corresponding
database query created a sample of 122 million objects with $u$, $g$, $r$, $i$,
and $z$-band {\it PSF} magnitudes and errors.

\subsubsection{Classification Model}

\begin{figure}
\begin{centering}
\includegraphics[width=1.0\columnwidth]{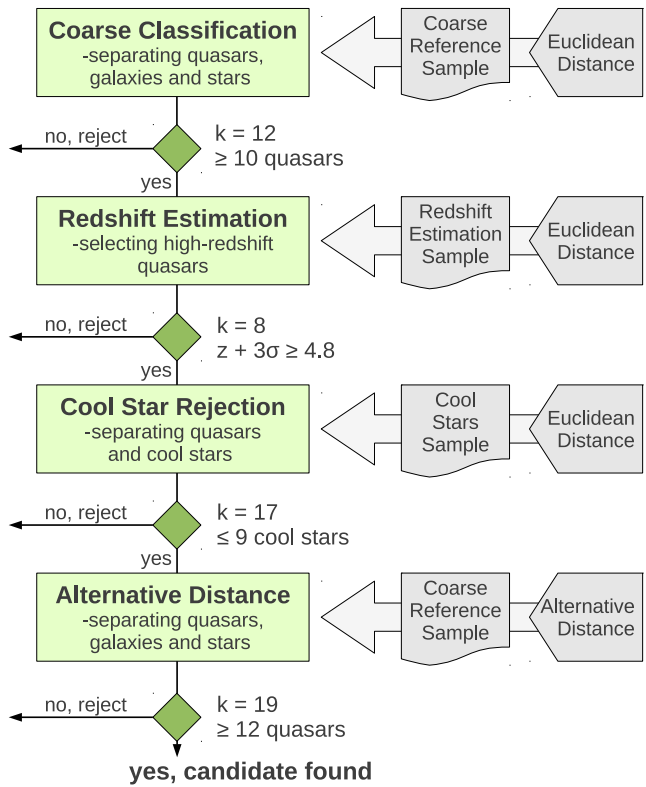}
\caption{Diagram of the different classifier stages. Both, the required
reference sample and the distance function which is used for the neighbourhood
search, are presented for every stage. In addition, the selection criteria
that have been applied are shown.}
  \label{selectionStages}
\end{centering}
\end{figure}

Similar to the redshift estimation approach presented in \mbox{Section
\ref{redshiftEstimation}}, the classification is done with the {\it kNN}
algorithm \citep[][]{hast:2009}. Based on the reference sample $S =
\{({\bf x}_1,t_1), \ldots, ({\bf x}_n,t_n)\} \subset {\mathbb R}^d \times T$,
the classification of the types $t_i$ of the $k$ nearest objects $N_k({\bf x})$
in the feature space is evaluated for each ${\bf x}$. The corresponding ratio $\hat
R_t({\bf x})$ reflects the number of the $k$ neighbouring reference
objects of ${\bf x}$ that are of a certain type $t \in T$ and is
defined as:

\begin{equation}
\hat R_{t}({\bf x})=\frac{1}{k}\sum
\limits_{{\bf x}_i \epsilon N_k({\bf x})} q_i
{\rm{\,with\,}} q_i = \begin{cases} 1, & t_i = t \\ 0, & \rm{otherwise} \end{cases}
\end{equation}

The quasar candidate selection is realised by combining the redshift estimator
and three classifiers (see \mbox{Figure \ref{selectionStages}}):

\begin{enumerate}
  \item The first classifier should realise a coarse quasar
pre-selection. This is achieved by using a more general reference set with
several object types. The Euclidean distance in the {\it PSF} magnitude feature
space is used to identify the {\it k} nearest neighbours.
  \item In the next step, the redshift is estimated and low-redshift objects are
  rejected (see \mbox{Section \ref{redshiftEstimation}}).
  \item In the third step, a classifier which rejects cool stars is used to
  decrease the contamination by stellar objects. This is realised by a comprehensive reference
sample which contains only cool stars and quasars. The same feature space as
for the coarse pre-selection is used.
  \item In the last step, an alternative distance measure $d$ is
used to run a classification with respect to the photometric errors. The first
part of this function reflects the similarity of two feature vectors
${\bf u}, {\bf v} \in {\mathbb R}^d$ with
respect to the measurement errors ${\bf {\Delta
u}}, {\bf {\Delta v}}$:

\begin{equation}
d({\bf u}, {\bf {\Delta u}}, {\bf v}, {\bf {\Delta
v}}) = \sum_{i=1}^{N} \frac{(u_i-v_i)^2}{\Delta u_i^2 + \Delta v_i^2} + (|\Delta
u_i| - |\Delta v_i| )^2
\end{equation}

The second part ensures that objects with similar errors become closer. When two
objects with severely deviating measurement errors are compared, the first
distance component decreases due to the dominant error term. This is compensated
by the second component.

\end{enumerate}

\begin{table*}
\begin{centering}
  \begin{tabular}{c|c|l}
  Name of Sample & Purpose & Sample Composition\\\hline\hline
   
  & & - all 1,258 ${\rm z} \geq 4.8$ high-redshift quasars\\
  & & - 1,000 randomly selected $2.0 \leq {\rm z} < 4.8$ medium-redshift quasars\\
   coarse & \raisebox{1.5ex}[-1.5ex]{coarse preselection,} & - 1,000 randomly
   selected galaxies\\
  & \raisebox{1.5ex}[-1.5ex]{used in stage 1 + 4} & - 1,000 randomly selected
  stars\\
  & & - 1,500 randomly selected cool stars\\\hline
  
  & rejection of cool stars, & - all 1,258 ${\rm z} \geq 4.8$ high-redshift
  quasars\\
  \raisebox{1.5ex}[-1.5ex]{cool stars} &
  used in stage 3 & - all spectroscopically determined cool stars\\\hline
  & & - 120 equal redshift bins between $1.0 \leq {\rm z} < 7.0$\\
  redshift & \raisebox{1.5ex}[-1.5ex]{estimation of redshift,} & \hspace{1ex}
  with a maximum of 10 quasars per bin\\
   & \raisebox{1.5ex}[-1.5ex]{used in stage 2}& - all ${\rm z} \geq 4.8$
   high-redshift quasars
   
  \end{tabular}
  \caption{The reference samples that are used for selecting
  quasar candidates and for estimating the redshifts.}
  \label{referenceSample}
\end{centering}
\end{table*}

\subsubsection{Reference Samples}

The classifiers require reference samples similar to the presented redshift
estimator sample. All object classifications which are used to create the
samples are based on spectroscopic observations (see \mbox{Table
\ref{referenceSample}}). The first sample was created to detect quasars and
represents five different types. The second table contains 10,928
objects and is used to reject cool stars. By using spectroscopically classified
objects and objects detected by the $i$-band dropout method as reference, the
applied sample selection criteria
\citep[][]{rich:2002,fan:2001,fan:2003,fan:2004,fan:2006} are reproduced by the
presented reference samples. Furthermore the objects selected as quasar
candidates by \citet{rich:2002} that turned out to be cool stars improve the
separation capabilities of the samples.

\subsubsection{Model Selection}
In each step of the selection process, objects that do not match the ratio
criteria are rejected. The ratios have been created on basis of the
{\it SDSS} objects with spectroscopic classifications. They have been optimised to
find as many high-z quasars as possible while simultaneously minimising the
contamination by other objects.

\begin{enumerate}
  \item For the coarse pre-selection step, a ratio of $10/12$ \mbox{high-z}
  quasars was determined.
  \item The redshift estimator is used to exclude objects with $\rm{z}\leq4.8$.
  The standard deviation that is calculated based on the {\it k} nearest
  neighbours is used to allow an undershooting of this $\rm{z}$ value by
  $3\sigma$.
  \item For the cool stars rejecting classifier a ratio of $8/17$ quasars was
empirically found to produce the best results. This means that ten or more cool
stars are required to reject a candidate.
  \item In the last selection step, the 19 nearest neighbours are inspected.
  Twelve of these neighbours must be high-z quasars to pass this step.
\end{enumerate}

\subsubsection{Post-Processing}
The objects that pass all four selection steps are written to a result file.
Instead of simple ratios it contains the types of the 20 nearest neighbours for
each step. This allows a post-processing to reduce the candidate list for
follow-up observations. The likelihood of a candidate being a high-z quasars can
be estimated by calculating different ratios afterwards. Thereby constraints can
be combined. These constraints can be separately specified for each of the
classifiers.

\subsection[Evaluation]{The Evaluation of the Selection Approach}

To evaluate the performance of the presented selection approach, all 1.2 million
{\it SDSS} objects with spectra have been photometrically analysed.

\subsubsection{Results}

Applying our approach for identifying ${\rm z}>4.8$ quasars to the 1.2
million SDSS objects with spectra, we obtained a list of 242 candidate high-z
quasars. We note that we explicitly excluded all such objects from neighbourhood
queries that were also used in one of the reference data sets for training the
model in order to prevent any biasing of our results. The spectra of all of
these 242 quasar candidates have been visually inspected subsequently
and their types have been checked. 147 of these candidates are quasars of the
high-z reference sample with $\rm{z}>4.0$. 75 of the known 147 quasars with
$\rm{z}>4.8$ are recovered by the photometric selection. The 18 quasars of the
high-z extension sample have been tested separately as they are not part of the
{\it SDSS} spectroscopic catalogue. 17 of these quasars can be detected with the
photometric selection approach. For these objects that have been found with an
$i$-band dropout method the coarse classifier calculates ratios of $\geq 19$ out
of 20. Of all 32,210 known cool stars, only 34 have been falsely classified as
quasars. The spectra of the remaining 61 objects are of other types or are not
assignable to a type.

\section{Results}
\label{results}

\begin{figure}
\begin{centering}
\includegraphics[width=0.9\columnwidth]{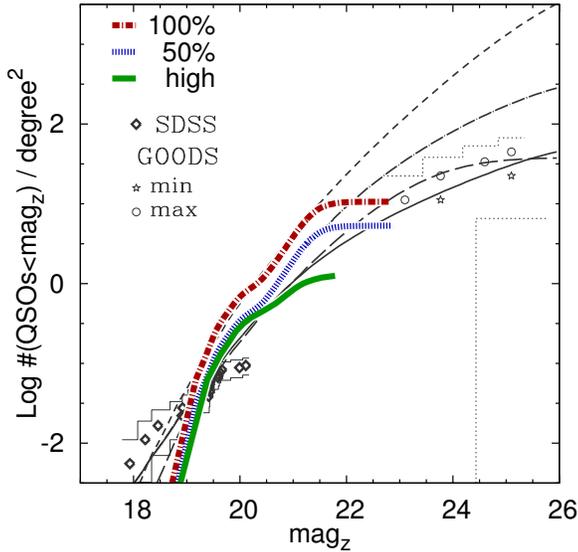}
\caption{Comparison of the space density plot of quasars ($4.0<\rm{z}<5.2$) from
\citet{chri:2004} and the parameters derived from the created candidate sample.
Circles and stars show the {\it GOODS} based estimates of \citet{chri:2004} while
diamonds are used for {\it SDSS} results. The tiny dashed lines represent the upper
and lower $1\sigma$ confidence level of the {\it GOODS} data. The models that have
been used by \citet{chri:2004} are: {\it PLE} (dot-dashed), {\it PDE} (continuous), {\it MIN}
(short-dashed) and {\it DEL} (long-dashed). The three coloured lines represent
the entire candidate sample (100 per cent), an assumed detection performance
(50 per cent) and the candidates with the highest \mbox{ratios $\hat R$}.
}
  \label{spaceDensity}
\end{centering}
\end{figure}

Finally, we used our approach to search the entire $DR6$ photometric
database for high-z QSOs. The $DR6$ database contains about 122\,million
objects down to $i$-band magnitudes of $\sim21.5$ which is more than two
magnitudes deeper than the limiting magnitude for SDSS spectroscopy. We
therefore anticipate to reveal a great number of formerly unconsidered QSOs with
${\rm z}>4.8$ to be potential targets for future spectroscopic follow-up.
The pre-selected 122 million objects have been processed with our approach and a
list was created.\footnote{As the execution of the individual classification steps
is not order-dependent they have been arranged by their processing speed.
This ensures that computing intensive steps are only executed when previous
classifications have been successful. Therefore, it can easily be parallelised
to increase the speed of processing lists. A single instance is able to process
1,000 objects in 4 -- 8\,s on a standard PC. For the creation of the final
candidate list the software was running on eight cores and processed the
122\,million photometric data sets in a day.} 
We obtained a list of 121,909 objects that match the specified ratio
criteria of the individual selection steps and are hence regarded as
\mbox{high-z} quasar candidates.
The sample contains redshift estimations for 82,358 candidates with
$\rm{z}<5.0$, 34,359 with $5.0\le\rm{z}<6.0$ and 5,193 with $\rm{z}\ge6.0$. The
detection performance was calculated on the spectroscopic sample to be about
$50$ per cent. 
A machine-readable version of the sample in comma separated value
format can be downloaded at MNRAS or directly at {\tt \small
http://www.astro.rub.de/polsterer/quasar-candidates.csv}. It contains the
original SDSS photometric ID, position, estimated redshift and its standard deviation,
the three ratios from the different stages of the classifiers and the PSF
magnitudes in all five SDSS bands. The redshift estimations have to be treated
carefully and should just be used as an indication. Especially the high-z area
is so underrepresented in the sample that a high standard deviation of the
redshift estimation indicates a possible outlier.

\vskip 3ex
\noindent
\framebox[\columnwidth]{
\begin{minipage}{0.95\columnwidth}
Under the assumption that the classifier performs
comparably on the photometric sample, $\approx60,000$ high-z quasars are part of
the candidate list.
\end{minipage}
}

\subsection{Comparison with Theoretical Models}
\citet{chri:2004} presented a space density for quasars with $4.0<\rm{z}<5.2$. A
separate classification run for this redshift range yielded a list of 102,825
candidates. In \mbox{Figure \ref{spaceDensity}} these candidates are plotted in
comparison with the results of \citet{chri:2004}. The red line is a cumulative
plot of all candidates while the blue line assumes a quasar detection
performance of about $50$ per cent. When the ratio of the first selection step
is set to the highest possible ratio (i.e. twelve quasars under the twelve nearest
neighbours) only 12,586 candidates remain. These candidates with the highest
probability of being a quasar are plotted as a green line.

The presented results are consistent with the results of \citet{chri:2004} and
the number of found quasar candidates fits well with the presented models. A
model which connects quasars and dark matter halos with a minimal set of
assumptions ({\it MIN}) is presented by \citet{haim:2001}. This model assumes:
(i) an un-evolved halo, (ii) a constant black hole to dark matter halo mass ratio
(iii) and a maximum accretion at the Eddington limit \citep[][]{chri:2004}. As
this model overpredicts high-z quasars \citep[][]{haim:1999}, it defines an
upper limit for the presented candidate selection approach. \citet{mona:2000} present
a model with a delayed quasar shining ({\it DEL}) in which the {\it AGN} activity starts
after the formation of the dark matter halo. In their model, {\it AGN}s which
are hosted in smaller halos are longer delayed than those {\it AGN}s in larger halos. This allows
brighter quasars to appear before the fainter ones. The pure luminosity
evolution ({\it PLE}) model (brighter objects in the past) and the pure density
evolution ({\it PDE}) model (higher object density in the past) are used in
\citet{chri:2004} to extrapolate the results of \citet{boyl:2000}. Both, the
high ratio results as well as the results of the predicted detection performance
fit well with these models. In {\it DR6} of the {\it SDSS} the 95 per cent
detection repeatability for point sources in the $z$-band is 20.5\,mag. For this reason the results with the
highest probability deviate from the model fits for higher $z$-band magnitudes.
The results of the other candidate lists fit well for $z$-band magnitudes below
21.5\,mag. In comparison to the quasars detected in the {\it SDSS}, an
appropriate amount of candidates can be found even for $z$-band magnitudes
fainter than 19.5\,mag (see \mbox{Figure \ref{spaceDensity}}).

\subsection{Spectroscopic Verification of Candidates}

\begin{figure}
\begin{centering}
\includegraphics[width=0.98\columnwidth]{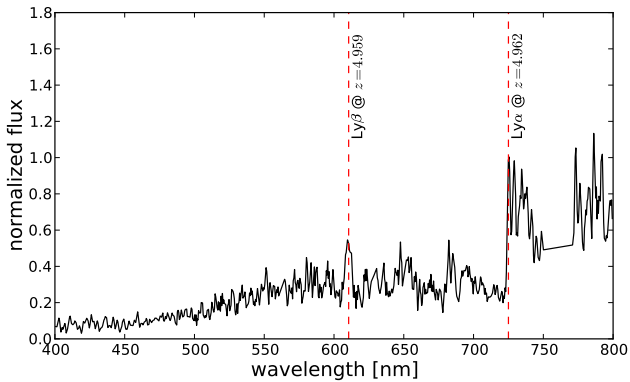}
\caption{Reduced spectrum of J1538+5032 (SDSS ID 588011218070144145), a
newly found quasar with a redshift of $\rm{z} \sim 5.0$. Taken with
SCORPIO@BTA.}
  \label{observationOne}
\end{centering}
\end{figure}

\begin{figure}
\begin{centering}
\includegraphics[width=1.0\columnwidth]{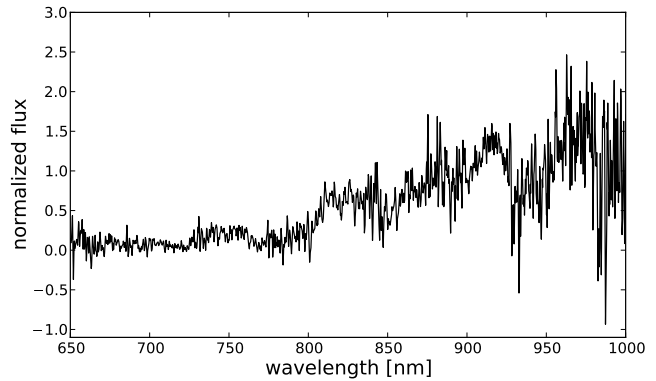}
\caption{Reduced spectrum of J1458+5436 (SDSS ID 588011216993125015), probably a
cool star. Taken with SCORPIO@BTA.}
  \label{observationTwo}
\end{centering}
\end{figure}

To finally get an impression of the overall performance of our quasar
identification approach, we randomly selected three candidates from our result
list for spectroscopic follow-up observations in order to confirm their quasar
nature and determine their redshift. This very small number of follow-up targets
is of course not sufficient to establish an empirical performance of our
algorithm in any statistically meaningful sense, but the identification of even
one high-z quasar would at least highlight the potential of data mining and
machine learning approaches for data-intensive astronomy and also demonstrate a
possible way to go with this new branch of astronomy: Since all outcomes of
follow-up spectroscopy are valuable for evaluating and optimising a kNN-based
classification approach (e.g. sources showing non-quasar spectra can be used to
enhance the training set for rejecting contaminating objects such as cool
stars), this method would greatly benefit from spectroscopy of a considerable
number ($>1000$) of candidate sources.

\subsubsection{Observations}

The observations were carried out in July 2011 with the
SCORPIO instrument at the 6-metre BTA telescope of the Special
Astrophysical Observatory on Mt. Pastukhova in southern Russia. A total
on-source time of 6.5 hours was spent on three objects that were
randomly chosen from our final candidate catalog, only applying a
magnitude cut of $i<22.0$\,mag in order to obtain reliable spectra in such a
short time. To optimise the efficiency of observations, we chose
low-resolution grisms GR300R and VPHG550R (a holographic grism)
alongside with a wide slit of $2\arcsec$, both resulting in a
theoretical spectral resolution of $\sim40\,\mathrm{\AA}$. Data
reduction was done using standard \texttt{IRAF} tasks, including the
removal of cosmic rays with \texttt{lacos-spec}. We did not apply an
absolute flux calibration since we are only interested in the general
shape of the spectrum and its redshift.

\subsubsection{Results}

From the three observed objects, the brightest one (SDSS ID
587735488676299501) with $i=20.30$\,mag turned out to be a cool star even in
the very first frames of observations. Therefore, we aborted the run for
this object and divided its observation time on the two remaining
sources with $i$-band magnitudes of 21.52\,mag (SDSS ID 588011218070144145)
and 21.45\,mag (SDSS ID 588011216993125015), respectively.
\mbox{Figure \ref{observationOne} and \ref{observationTwo}} show the final
wavelangth-calibrated 1D spectra of the objects. As one can clearly see, we have
one detection of a quasar with a very pronounced Lyman break at
$7,249\,\mathrm{\AA}$ and a broad Lyman $\beta$ emission line at
$6,125\,\mathrm{\AA}$ that correspond to a redshift of $\rm{z} \sim 5.0$,
regarding errors of the wavelength calibration due to the known flexure of the
CCD chip of SCORPIO. The other spectrum is not as easy to classify, therefore we
only state that this is not a quasar but most likely another cool star. However,
there are classes of objects known, which harbour an {\it AGN}
without showing distinct signs of {\it AGN}-activity in an optical spectrum
\citep{zinn:2011}. Therefore we can not make definite statements on the nature
of the sources which do not show any signs of {\it AGN}-activity in their
spectra.

\subsubsection{Outlook}

We were only able to observe three objects from our final
candidate catalog comprising more than 120,000 objects. Since
spectroscopy of such faint sources is very time consuming\footnote{Even at six
or eight metre class telescopes several hours are required to do follow-up
observations.} we cannot draw any significant statistical conclusions from these
observations regarding the fraction of actual quasars in our catalog.
Nevertheless, we point out, that in combination with the excellent agreement
with the high-z quasar number densities found in the previous section, the newly
identified $\rm{z}\sim 5.0$ quasar is a very promising result. To finally obtain
spectra of a substantial number of objects allowing for reliable statistical
analysis of our selection approach, many more nights at big telescopes have to
be spent.

\section{Conclusions}
\label{conclusions}

The presented quasar candidate selection approach is highly dependent on the
coverage of the feature space induced by the reference samples. Efficient data
structures are mandatory to provide good scanning performance when using large
reference samples. Except for the last step the {\it kNN} retrieval of all other
classifiers is accelerated by $k$-d-trees \citep[][]{bent:1975}. With the
availability of larger spectroscopic surveys and larger sets of known high-z
quasars as reference, better photometric selection results will be achievable.

Even though not all of the known quasars are recovered by the presented
approach, the resulting candidates have higher probabilities to be a quasar
than those found with other approaches. Follow-up observations will help to
determine the real detection performance of the candidate selection. The objects
that are found not to be quasars will directly improve the reference sets and,
thus, the overall detection performance.

\section*{Acknowledgments}

We thank Serguei N. Dodonov of the Special Astrophysical Observatory for
carrying out the follow-up observations on SCORPIO.

We thank Ralf-J\"urgen Dettmar, Dominik. J. Bomans, and Wolfhard Schlosser for
the helpful discussion on the topic of quasars and statistics.

This research has made use of the NASA/IPAC Extragalactic Database (NED) which is operated by the Jet Propulsion
Laboratory, California Institute of Technology, under contract with the National Aeronautics and Space Administration.

This research has made use of NASA's Astrophysics Data System Bibliographic Services

Based on data of Sloan Digital Sky Survey ({\it SDSS}). Funding for the {\it SDSS} and {\it SDSS}-II has been provided by the Alfred P.
Sloan Foundation, the Participating Institutions, the National Science Foundation, the U.S. Department of Energy, the
National Aeronautics and Space Administration, the Japanese Monbukagakusho, and the Max Planck Society, and the Higher
Education Funding Council for England. The {\it SDSS} Web site is http://www.sdss.org/. The {\it SDSS} is managed by the
Astrophysical Research Consortium (ARC) for the Participating Institutions. The Participating Institutions are the
American Museum of Natural History, Astrophysical Institute Potsdam, University of Basel, University of Cambridge, Case
Western Reserve University, The University of Chicago, Drexel University, Fermilab, the Institute for Advanced Study,
the Japan Participation Group, The Johns Hopkins University, the Joint Institute for Nuclear Astrophysics, the Kavli
Institute for Particle Astrophysics and Cosmology, the Korean Scientist Group, the Chinese Academy of Sciences
(LAMOST), Los Alamos National Laboratory, the Max-Planck-Institute for Astronomy (MPIA), the Max-Planck-Institute for
Astrophysics (MPA), New Mexico State University, Ohio State University, University of Pittsburgh, University of
Portsmouth, Princeton University, the United States Naval Observatory, and the University of Washington.

This research has made use of Aladin and Topcat.

\bibliographystyle{mn2e}
\bibliography{bib}


\bsp

\label{lastpage}

\end{document}